# Relativistic Spin-Lattice Interaction Compatible with Discrete Translation Symmetry in Solids

Bumseop Kim[1,2], Noejung Park[2,3*] and Kyoung-Whan Kim[4*]

[1]*Department of Chemistry, University of Pennsylvania, Philadelphia, PA 19104, United States*

[2]*Department of Physics, Graduate School of Semiconductor Materials and Devices Engineering, Ulsan National Institute of Science and Technology, Ulsan 44919, Korea*

[3]*Max Plank Institute for the Structure and Dynamics of Matter, Hamburg 22761, Germany*

[4]*Department of Physics, Yonsei University, Seoul 03722, Korea*

*corresponding author e-mail

noejung@unist.ac.kr; kwkim@yonsei.ac.kr

**Abstract**

Recent interest in orbital angular momentum has led to a rapid expansion of research on spin-orbit coupling effects in solids, while also highlighting significant technical challenges. The breaking of rotational symmetry renders the orbital angular momentum operator ill-defined, causing conceptual and computational issues in describing orbital motion. To address these issues, here we propose an alternative framework. Based on the Bloch representation of the full relativistic interaction, we derive a field that directly couples to electron spins while preserving discrete translational symmetry, thereby eliminating the need for the position operator. Our approach is fully compatible with existing first-principles computational frameworks for both static and time-dependent density functional theory. We demonstrate that this method offers a more effective description of the Edelstein and spin Hall effects compared to conventional orbital angular momentum formalisms.

## Introduction

Spin-orbit coupling (SOC) is an important element for various phenomena in modern condensed matter physics. It not only gives rise to fascinating equilibrium properties, such as spin-momentum locking [1-5], non-trivial topology [6,7], and anti-symmetric exchange interactions [8,9], but also leads to exotic transport phenomena, including the spin and anomalous Hall effects [10-12], the Edelstein effect [13,14], and spin-orbit torque [15-17]. Furthermore, the SOC-driven spin dynamics has been discussed in the context of next-generation device applications [18-20], highlighting the importance of comprehensive calculation of SOC effects.

The recent rebirth [21] of orbitronics [22,23] has brought the concept of orbital angular momentum (OAM) as a central tool for understanding SOC phenomena. For example, the intrinsic spin Hall effect in centrosymmetric normal metals is now interpreted as the spin counterpart of the orbital Hall effect [21,23,24]. This has led to the exploration of various orbital-related phenomena, such as orbital torque [25-27], the orbital Edelstein effect [28,29], orbital angular position [30,31], orbital pumping [31,32], and orbital diffusion [33]. Moreover, experimental demonstrations of the orbital Hall effect [34,35] have garnered significant attention.

Despite these practical advancements, any attempt to attain OAM in solids encounters a fundamental conceptual challenge: in the absence of continuous rotational symmetry, OAM is inherently ill-defined. The most conventional definition of OAM, $\hat{\mathbf{L}} = \hat{\mathbf{r}} \times \hat{\mathbf{p}}$, relies on the position operator $\hat{\mathbf{r}}$, which is not well-defined in translationally symmetric systems. Consequently, the matrix element of $\hat{\mathbf{r}}$ between Bloch states leads to divergences near degeneracies [36-39]. To circumvent this issue, the atom-centered approximation (ACA) is commonly employed, wherein the Wannier function is expanded in terms of spherical harmonics to construct the intra-atomic contribution to the OAM operator [25,40,41]. However, it is widely recognized that the inter-atomic contributions are not negligible [42,43]. Moreover, the nonlocality of metallic systems and the ambiguity of Wannier functions hinder the versatility of this method, particularly for extended quantities such as orbital current.

Previous efforts to solve these issues include considering finite systems [44] or treating inter-atomic contributions separately [42]. The former is unsuitable for studying non-equilibrium angular momentum flow, which has garnered increasing interest [21], while the latter does not reproduce results consistent with the modern theory of orbital magnetism [44-47] and introduces ambiguities in interpretation [48]. From a computational perspective, the position operator introduces terms proportional to the inverse of the energy difference between two states [42,43], resulting in numerical instabilities, particularly in nonequilibrium [49]. These conceptual and technical difficulties originate from the use of the position operator when defining the OAM in solids.

While the OAM is a convenient tool in atomic physics, it is less suitable for solids, where the discrete translational symmetry governs the physics. In this Letter, we propose a framework that eliminates the need for the position operator, and introduces an alternative operator to OAM for describing SOC phenomena [50]. By projecting the full relativistic interaction into the Bloch basis, we derive the relativistic spin-lattice interaction (SLI) field, denoted by $\mathbf{\Lambda}$, in a form fully compatible with existing first-principles computational techniques. Our first-principles calculation for its equilibrium textures, and the Edelstein/Hall effects associated with $\mathbf{\Lambda}$, and time-dependent responses show that $\mathbf{\Lambda}$ effectively describes SOC phenomena while overcoming the limitations of the conventional OAM operator.

## Bloch representation of the relativistic SLI

Our starting point is the spin component of the relativistic interaction, commonly referred to as SOC.

$$\hat{H}_{\mathrm{rel}} = \frac{\hbar}{4m_e^2c^2}\hat{\boldsymbol{\sigma}} \cdot \left(\nabla\hat{V} \times \hat{\mathbf{p}}\right), \tag{1}$$

where $m_e$ is the electron mass, $c$ is the speed of light, $\hat{\boldsymbol{\sigma}}$ consists of the Pauli matrices, $\hat{\mathbf{p}}$ is the momentum operator, and $\hat{V}$ is the full lattice potential. In the conventional OAM formalism, $\hat{V}$ is often replaced by the sum of local potentials, such as $Ze^2/4\pi\epsilon_0 r$, rewriting Eq. (1) as a sum of terms $\propto (1/r^3)\hat{\mathbf{S}} \cdot \hat{\mathbf{L}}$ where $\hat{\mathbf{S}} = \hbar\hat{\boldsymbol{\sigma}}/2$ and $\hat{\mathbf{L}} = \hat{\mathbf{r}} \times \hat{\mathbf{p}}$ are the spin and OAM operators, respectively. This approach has several issues: (i) It relies on a local approximation for $\hat{V}$, unsatisfactory in metallic systems with delocalized electronic states. (ii) It involves the position operator $\hat{\mathbf{r}}$, whose subtlety was discussed in the introduction. (iii) Each Bloch state $|\mathbf{k}n\rangle$ can have different coupling strength $\xi_{\mathbf{k}n} \propto \langle \mathbf{k}n|1/r^3|\mathbf{k}n\rangle$, complicating the interpretation of the orbital-to-spin conversion (and vice versa). It makes the spin coupled with state-dependent quantity ($\sum_{\mathbf{k}n} \xi_{\mathbf{k}n}\langle \mathbf{k}n|\mathbf{L}|\mathbf{k}n\rangle$) rather than directly to the total OAM $\sum_{\mathbf{k}n}\langle \mathbf{k}n|\mathbf{L}|\mathbf{k}n\rangle$. A previous study [51] disproved the correlation between the spin Hall conductivity (SHC) and orbital Hall conductivity (OHC). These issues arise from the introduction of $\mathbf{L}$, which requires the inclusion of $1/r^3$ and its associated complications.

We thus define the field of relativistic SLI in the following form.

$$\hat{\boldsymbol{\Lambda}} = \eta\nabla\hat{V} \times \hat{\mathbf{p}}, \tag{2}$$

where $\eta = m_e a_0^4/\hbar^2 = 52.59\ (\mathrm{nm}^2/m_e c^2)$ and $a_0$ is the Bohr radius. In terms of this definition, Eq. (1) can be rewritten as $H_{\mathrm{rel}} = \xi_{\mathrm{rel}}\hat{\mathbf{S}} \cdot \hat{\boldsymbol{\Lambda}}$, where $\xi_{\mathrm{rel}} = \alpha_{\mathrm{FSC}}^2/2m_e a_0^2 = 0.7245\ (\mathrm{meV}/\hbar^2)$ is a *universal* constant and $\alpha_{\mathrm{FSC}}$ is the fine-structure constant. The physical meaning of the relativistic SLI is the effective field experienced by electron spins arising from the relativistic corrections, which originate from the electric dipole moment generated by a moving spin [52] and its Coulomb interaction with the lattice. $\hat{\mathbf{\Lambda}}$ is on the order of $\hbar$ when $\nabla V \sim e^2/4\pi\epsilon_0 a_0^2$ and $\mathbf{p} \sim \hbar/a_0$.

We highlight the advantage of introducing $\mathbf{\Lambda}$. First, $\hat{V}$ represents the full lattice-periodic potential and does not rely on any local approximation or Wannierization, thereby resolving the issue (i) above. Second, $\hat{\mathbf{\Lambda}}$ includes only the periodic part of the position operator, confined within the unit cell, thus avoiding the conceptual and technical difficulties associated with the divergence of the position operator, thereby addressing issue (ii). Third, since $\eta$ and $\xi_{\text{rel}}$ are universal constants, the spin angular momentum couples directly to $\hat{\mathbf{\Lambda}}$ even after summation over electronic states, resolving issue (iii). Most importantly, its periodicity guarantees full compatibility with the symmetry of solids, and the Bloch representation to be used without any conceptual ambiguity.

To derive the matrix elements of $\hat{\mathbf{\Lambda}}$ in the Bloch basis, we considering the full lattice Hamiltonian $\hat{H} = \hat{\mathbf{p}}^2/2m_e + \hat{V} + \hat{H}_{\text{rel}}$, where the gradient of $\hat{V}$ can be expressed as $\nabla\hat{V} = \nabla\hat{H} - \nabla\hat{H}_{\text{rel}} = (i/\hbar)\left([\hat{\mathbf{p}}, \hat{H}] - [\hat{\mathbf{p}}, \hat{H}_{\text{rel}}]\right)$. Feeding this back to Eq. (1), we obtain the following recursive relation for $\hat{H}_{\text{rel}}$.

$$\hat{H}_{\text{rel}} = \frac{i}{4m_e^2c^2}\hat{\boldsymbol{\sigma}} \cdot \left([\hat{\mathbf{p}}, \hat{H}] \times \hat{\mathbf{p}}\right) - \frac{i}{4m_e^2c^2}\hat{\boldsymbol{\sigma}} \cdot \left([\hat{\mathbf{p}}, \hat{H}_{\text{rel}}] \times \hat{\mathbf{p}}\right). \tag{3}$$

If SOC is weak, the leading-order contribution to $1/c^2$ is given by the first term: $\hat{H}_{\text{rel}} = \xi_{\text{rel}}\hat{\mathbf{S}} \cdot \hat{\mathbf{\Lambda}}$, where $\hat{\Lambda}_\mu = (\eta/i\hbar)\left(\hat{\mathbf{p}} \times [\hat{\mathbf{p}}, \hat{H}]\right)_\mu = (\eta/2i\hbar)\epsilon_{\mu\nu\lambda}\{\hat{p}_\nu, [\hat{p}_\lambda, \hat{H}]\}$. The matrix element of $\hat{\mathbf{\Lambda}}$ in the Bloch basis is then

$$\langle u_{n\mathbf{k}}|\widehat{\mathbf{\Lambda}}|u_{m\mathbf{k}}\rangle = \frac{\eta}{2i\hbar}\langle u_{n\mathbf{k}}|\widehat{\mathbf{p}} \times (E_{n\mathbf{k}} + E_{m\mathbf{k}} - 2H_{\mathbf{k}})\widehat{\mathbf{p}}|u_{m\mathbf{k}}\rangle, \tag{4a}$$

where $|u_{n\mathbf{k}}\rangle = e^{-i\mathbf{k}\cdot\hat{\mathbf{r}}}|\psi_{n\mathbf{k}}\rangle$ is the cell-periodic part of the Bloch eigenstate $|\psi_{n\mathbf{k}}\rangle$ with the energy eigenvalue $E_{n\mathbf{k}}$, and $\widehat{H}_{\mathbf{k}} = e^{-i\mathbf{k}\cdot\hat{\mathbf{r}}}\widehat{H}e^{i\mathbf{k}\cdot\hat{\mathbf{r}}}$ is the reduced Hamiltonian in the **k** block. The momentum operator acting on the Bloch basis is $\widehat{\mathbf{p}} = \hbar\mathbf{k} - i\hbar\nabla_{\mathbf{r}}$. Another expression can be obtained by using $\hat{\mathbf{v}} = (1/i\hbar)[\hat{\mathbf{r}}, \widehat{H}] = \widehat{\mathbf{p}}/m_e + (\hbar/4m^2c^2)\nabla\widehat{V} \times \widehat{\boldsymbol{\sigma}}$. Since the second term can be neglected due to its higher-order SOC contribution, it gives an alternative expression for Eq. (4a) as

$$\langle u_{n\mathbf{k}}|\widehat{\mathbf{\Lambda}}|u_{m\mathbf{k}}\rangle = \frac{m_e^2\eta}{2i\hbar}\langle u_{n\mathbf{k}}|\hat{\mathbf{v}} \times (E_{n\mathbf{k}} + E_{m\mathbf{k}} - 2\widehat{H}_{\mathbf{k}})\hat{\mathbf{v}}|u_{m\mathbf{k}}\rangle, \tag{4b}$$

where the velocity operator acting on the Bloch basis is $\mathbf{v} = (1/\hbar)\,\partial_{\boldsymbol{k}}H_{\mathbf{k}}$. Equation (4), which is a central result of this Letter, can be computed using information readily available from first-principles calculations. Possible higher-order SOC effects are discussed in Supplemental Materials [53].

Several important remarks follow. First, Eqs. (3) and (4) enable the calculation of the relativistic SLI without considering conventional forms of SOC, to arbitrary order in $1/c^2$. For nonmagnetic materials, the spin degree of freedom can be turned off, and the SLI can be computed with a substantially lower computational cost. Second, a comparison of Eq. (4a) with the orbital magnetization operator in Ref. [42] shows that the covariant gradient $|\partial_{\mathbf{k}}u_{\mathrm{n}\mathbf{k}}\rangle = (1 - |u_{n\mathbf{k}}\rangle\langle u_{n\mathbf{k}}|)(\nabla_{\mathbf{k}}|u_{n\mathbf{k}}\rangle)$ is replaced by the momentum operator. Another comparison can be made with Eq. (4b), which is equivalent to $\langle u_{n\mathbf{k}}|\mathbf{\Lambda}|u_{m\mathbf{k}}\rangle \propto -\sum_q(E_{q\mathbf{k}} - E_{n\mathbf{k}} + E_{q\mathbf{k}} - E_{m\mathbf{k}})\langle u_{n\boldsymbol{k}}|\mathbf{v}|u_{q\boldsymbol{k}}\rangle \times \langle u_{q\boldsymbol{k}}|\mathbf{v}|u_{m\boldsymbol{k}}\rangle$. This computational procedure resembles that of the orbital magnetization operator when $(E_{q\mathbf{k}} - E_{n\mathbf{k}}) + (E_{q\mathbf{k}} - E_{m\mathbf{k}})$ is replaced by $(E_{q\mathbf{k}} - E_{n\mathbf{k}})^{-1} + (E_{q\mathbf{k}} - E_{m\mathbf{k}})^{-1}$. This indicates that our theory does not suffer from the aforementioned technical difficulties rooted in the energy differences in the denominator. Third, since the off-diagonal element of the velocity operator is intimately connected to the $\mathbf{k}$-space gradient of $|u_{n\mathbf{k}}\rangle$, the nontrivial topology arising from it whould affect the SLI field. To demonstrate this explicitly, we have considered the topologically nontrivial material $Sn_2H$ (Supplementary Note 3 [53]) to show its correlation with the Berry curvature. Further investigation along this direction would be of significant interest.

## First-principles calculations: Equilibrium textures

We perform the first-principles calculations of Eq. (4a) for exemplary three-dimensional (3D), two-dimensional (2D), and one-dimensional (1D) materials. The computational details are shown in Supplemental Materials [53]. We first apply our theory to insulating or semiconducting materials, where the electronic wave functions are well localized near the atomic centers, and examine whether our results align with the intra-atomic OAM in ACA by comparing $\mathbf{\Lambda}^{n,\mathbf{k}} = \langle u_{n,\mathbf{k}}|\hat{\mathbf{\Lambda}}|u_{n,\mathbf{k}}\rangle$ and $\mathbf{L}^{n,\mathbf{k}} = \langle u_{n,\mathbf{k}}|\hat{\mathbf{L}}_{\mathrm{ACA}}|u_{n,\mathbf{k}}\rangle$. Here, $\hat{\mathbf{L}}_{\mathrm{ACA}}$ is the OAM operator in the ACA with the maximally localized Wannier function [53]. To avoid orbital quenching, we choose inversion-asymmetric systems, such as GaAs (3D), *h*-BN monolayer (2D), and Se chain (1D), depicted in Figs. 1(a)-(c). The results in Fig. 1 are obtained without SOC; however, its inclusion does not affect our conclusions [53]. Computed electronic structures in Figs. 1(d)-1(f) agree well with previous reports [54-56] and the momentum-space profiles of $\Lambda_z^{n,\mathbf{k}}$ and $L_z^{n,\mathbf{k}}$ exhibit very similar trends.

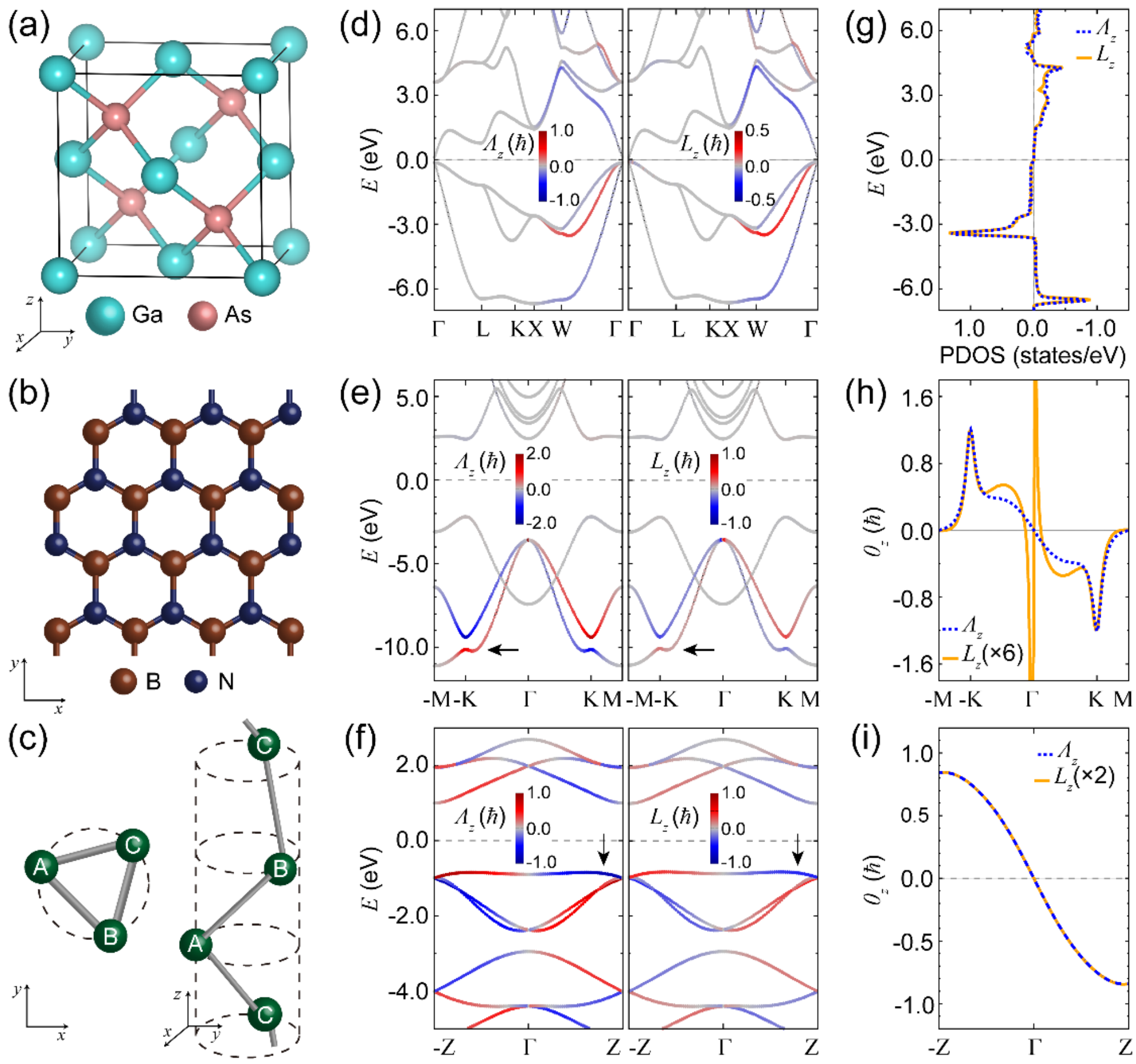

Fig. 1 (Color online) (a)-(c) Atomic structures of 3D GaAs (a), 2D *h*-BN monolayer (b), and 1D helical Se chain (c). (d)-(f) Calculated band structures of GaAs (d), *h*-BN monolayer (e), and Se chain (f) with momentum-resolved $\Lambda_z^{\mathbf{k}}$ (left panels) and $L_z^{\mathbf{k}}$ within the ACA (right panels). (g) $\Lambda_z$- and $L_z$-weighted partial density of states for GaAs. (h) and (i) Momentum-resolved $\Lambda_z^{\mathbf{k}}$ and $L_z^{\mathbf{k}}$ for the *h*-BN monolayer (h) and the Se chain (i).

To be more quantitative, we computed the $\Lambda_z^{n,\mathbf{k}}$- and $L_z^{n,\mathbf{k}}$-weighted partial density of states (PDOS) for GaAs [53]. In Fig. 1(g), these exhibit remarkable similarity, except for tiny discrepancies above the Fermi level, attributed to orbital hybridization (Fig. S1 [53]). For the *h*-BN monolayer and the Se chain, we compare $\Lambda_z^{n,\mathbf{k}}$ and $L_z^{n,\mathbf{k}}$ of a specific band, highlighted by black arrows in Figs. 1(e) and 1(f): Figures 1(h) and 1(i) demonstrate that the two values are quite well overlap, besides the overall scale. This consistency persists over different bands and almost unaffected by the inclusion of SOC (Figs. S2 and S3 [53]). Unlike $L_z$, which exhibits divergence near degeneracy, $\Lambda_z$ remains well-behaved over the entire range [Fig. 1(h)], indicating its superior numerical stability in describing the relativistic Hamiltonian.

In cases of metallic systems with delocalized charge distributions, $\mathbf{\Lambda}$ and OAM may display quantitative differences, offering an opportunity to determine which quantity is better associated with spin. To explore this further, we consider the $BiAg_2$ monolayer, which exhibits both orbital-Rashba and spin-Rashba effects [57-60] due to the *z*-directional displacement $\Delta d$ in Fig. 2(a). The calculated electronic structures in Fig. 2(b) reveal the Rashba-type orbital splitting, even in the absence of SOC [61,62], and the spin-Rashba effect follows upon the inclusion of SOC. The significant orbital splitting guarantees the full recovery of our *ab initio* electronic structure using maximally localized Wannier functions (Fig. S4 [53]).

We consider the Edelstein effect arising from the spin/orbital textures. When an electric field (along *x*) is applied, the shifted Fermi surface gives rise to nonzero spin/orbital densities (along *y*), as depicted in the inset of Fig. 2(c). Although this is a nonequilibrium phenomenon, it effectively reflects the equilibrium $\mathbf{k}$-space texture. The Edelstein effects associated with spin, orbital, and $\mathbf{\Lambda}$ are calculated by $\theta_y(E) = \sum_{n,\mathbf{k}} f^{(1)}\left(E - E_{n,\boldsymbol{k}-\Delta k_x\hat{x}}\right)\langle u_{n,\mathbf{k}}|\hat{\theta}_y|u_{n,\mathbf{k}}\rangle$ where $E$ is the Fermi level, $\hat{\theta}_{\boldsymbol{y}} = \hat{\Lambda}_y, \hat{S}_y, \hat{L}_y$, and $f^{(1)}\left(E - E_{n,\boldsymbol{k}-\Delta k_x\hat{x}}\right)$ refers to the shifted Fermi-Dirac distribution. We consider both versions of OAM, $\hat{L}_y(\mathrm{ACA})$ and $\hat{L}_y(\mathrm{Mod})$, calculated by the ACA and the modern theory in Ref. [44,63], respectively. $\Delta k_{\boldsymbol{x}} = 0.008$ Å$^{-1}$ (corresponding to 1 % of the Brillouin zone) and a thermal broadening of 0.002 Ry (corresponding to 316 K) in the Fermi-Dirac distribution are used, but the results are largely insensitive to thermal broadening factors within a reasonable range (Fig. S5 [53]). The SOC is turned off for computing $\mathbf{\Lambda}$ and $\mathbf{L}$, and turned on for $\mathbf{S}$ as the spin texture does not exist without SOC.

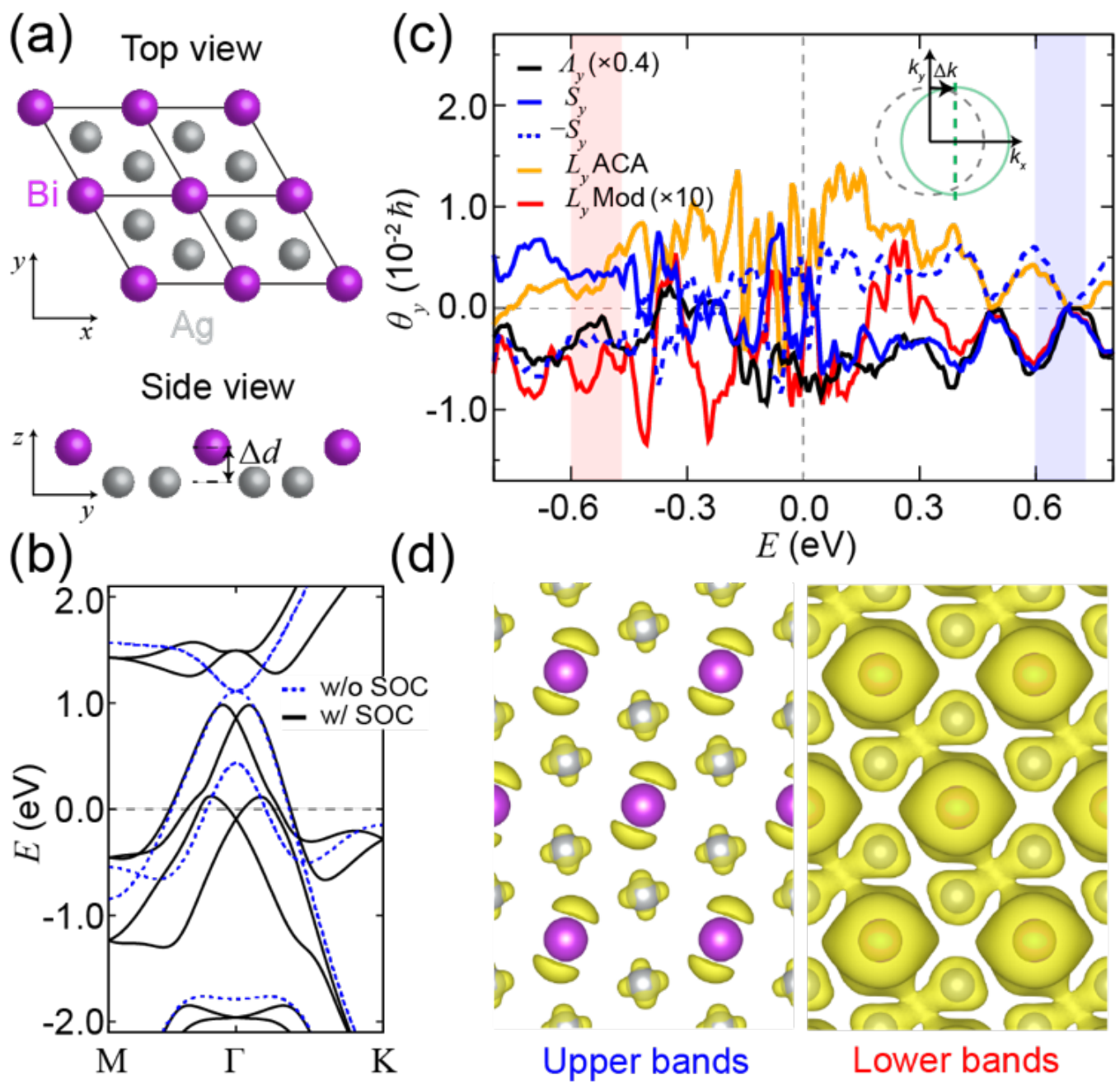

Fig. 2 (Color online) (a) Top and side views of a $BiAg_2$ monolayer. Displacement of Bi atoms from Ag layer is denoted as $\Delta d$. (b) Calculated band structure of the $BiAg_2$ monolayer with and without SOC. (c) The Edelstein components calculated by $\mathbf{\Lambda}$, spin, and OAM approximated by ACA, and Mod as a function of the Fermi level. The inset indicates schematic drawing of shifted Fermi surface as a response to electric field. (d) Real-space representation of the charge density of the upper and lower energy bands [blue and red in (c), respectively].

Figure 2(c) shows the results for $\theta_y(E)$. It reveals that the behavior of $\hat{\Lambda}_y$ (black line) resembles that of $\hat{S}_y$ (blue solid line) for $E > 0$ and that of $-\hat{S}_y$ (blue dashed line) for $E < 0$. The sign dependence arises from the fact that $\mathbf{\Lambda}$ and $\mathbf{S}$ are parallel (antiparallel) to each other in the upper (lower) band (Figs. S5 and S6 [53]). On the other hand, the behaviors of the calculated OAM [$L_y$ ACA and $L_y$ Mod in Fig. 2(c)] qualitatively differ from those of spin, except for $E > 0.3$ eV. The resemblance for $E > 0.3$ eV is attributable to the localized natures of the electronic states in the upper bands [left panel in Fig. 2(d)]. However, neither version of OAM mimics the behavior of $\mathbf{S}$ when the electronic states are delocalized [right panel in Fig. 2(d)]. These observations suggest that $\mathbf{\Lambda}$ describes relativistic spin phenomena more effectively than OAM in metallic systems with delocalized wave functions. Similar results were obtained for some topologically nontrivial materials (Supplementary Note 3 [53]).

## First-principles calculations: Hall conductivities

A crucial issue in understanding spin phenomena and OAM is the lack of numerical correlation between SHC and OHC [21,51]. We now focus on examining whether $\mathbf{\Lambda}$ provides a better description for such nonequilibrium phenomena. We choose bulk Pt, which exhibits large intrinsic spin and orbital Hall effects at room temperature [40,42,64,65], and calculate the Hall conductivities for $\mathbf{\Lambda}$, spin, and orbital using the Kubo formula [53]. The Wannier-interpolated band structure used for ACA accurately reproduces the *ab initio* band structure [Fig. 3(a)], aligning well with prior studies [40,64]. As shown in Fig. 3(b) (green line), the SHC reaches a maximum value of approximately $2300(\hbar/e)\,\mathrm{S/cm}$ near $E = -5$ eV, while the OHC exceeds this with values of $8000(\hbar/e)\,\mathrm{S/cm}$ and $5500(\hbar/e)\,\mathrm{S/cm}$ for ACA (orange line) and Mod (red line), respectively, near $E = -1$ eV, in good agreement with previous results [40,42,64]. The Hall conductivity of $\mathbf{\Lambda}$ (blue line) well correlated with the SHC over wide range of energy, which largely deviates from those of OHC. Furthermore, $\mathbf{\Lambda}$ offers a substantial computational cost advantage by enabling the examination of spin behavior without requiring spinor wave functions. Similar results were obtained for $Bi_2Se_3$ (Supplementary Note 3 [53]).

Hall conductivities are defined in nonequilibrium steady states. In Appendix A, we present our results from real-time time-dependent density-functional theory (rt-TDDFT) calculations which capture the fully dynamic responses of leading to consistent conclusions.

These findings validate the robustness of our results and suggest promising directions for future developments based on dynamical simulations.

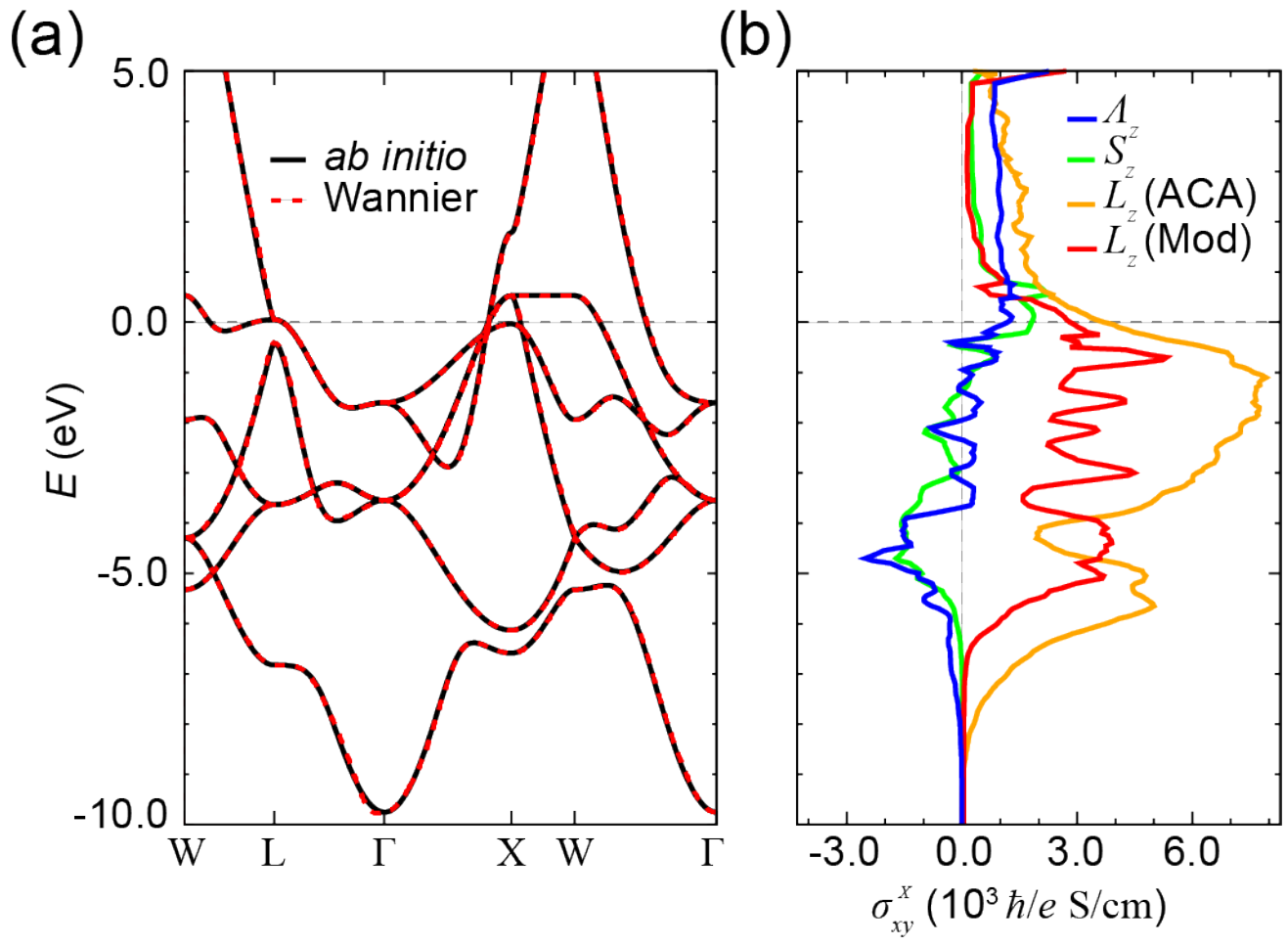

Fig. 3 (Color online) (a) Calculated band structure of bulk Pt by the *ab initio* (black solid line) and the maximally localized Wannier functions (red dotted line). (b) Hall conductivities of **Λ** (blue), spin (green), and OAM calculated by ACA (ACA, orange) and that by the modern theory (Mod, red).

## Discussion and summary

In this work, we derive a Bloch representation of relativistic spin-lattice interaction Hamiltonian, denoted by **Λ**, which can be directly implemented in standard first-principles band structure calculations methods. This provides an alternative but far superior definition of the operator compared to existing treatments of OAM, as it is free from conceptual and computational complexities rooted in the improper use of the position operator. Through first-principles calculations of both static and dynamical properties, we demonstrate that **Λ** can be obtained with enhanced numerical stability and reduced computational burden, and that it exhibits improved compatibility with spin angular momentum. We suggest this operator would be particularly meaningful for dynamical states of spins, with examples including orbital-to-spin conversion [21,23] and orbital torque on ferromagnets [26,27]. While the direct measurement of $\boldsymbol{\Lambda}$ remains challenging, we anticipate that its implications will motivate future efforts to identify experimental signatures of relativistic spin–lattice interactions. In particular, disentangling spin and orbital contributions in measurable quantities and exploring material platforms with strong SLI effects will be important directions for future research.

## Acknowledgement

The authors acknowledge D. Go, H. Jin, C.-J. Kang, B. H. Kim, K.-M. Kim, J. Kim, H.-W. Lee, K.-J. Lee, J. H. Oh, P. Oppeneer, Y.-W. Son, and G. Vignale for fruitful discussions. This work was supported by the National Research Foundation of Korea (NRF) grant funded by the Korea government (MSIT) (RS-2024-00334933, RS-2024-00410027, RS-2023-00257666, RS-2023-00218799, RS-2023-00208825), Yonsei University (2025-22-0089), and Samsung Electronics Co. through Industry-University Cooperation Project (IO221012-02835-01).

## End Matter

### Appendix A. Time-dependent calculations for optical responses

We here examine the dynamic properties of $\mathbf{\Lambda}$ (together with spin and OAM) in time-dependent simulations beyond the steady-state regime. We incorporated the effects of incident light using rt-TDDFT calculations (detailed information is in Supplemental Materials [53]). Since the Wannierization is computationally challenging in real-time calculations, we test only the OAM given by the modern theory.

Figure 4(a) presents the schematic drawing of rt-TDDFT calculations for the $BiAg_2$ monolayer. We applied an oscillating field along the *x*-direction (linearly polarized light), with an intensity of $0.001\ \mathrm{V/Å}$ and a frequency of $\hbar\omega = 1.0$ eV. As shown in Figure 4(b), the AC Edelstein effect along the *y* direction revealed that both spin and $\mathbf{\Lambda}$ converge to approximately $0.008\,\hbar$ within around 100 fs, while the orbital dynamics continued to increase, reaching about $0.02\,\hbar$ during the same period. This disparity can be attributed to carrier dynamics: after around 50 fs, no further changes occur in the spin-splitting states, whereas changes persist in the orbital-splitting states (Figs. S8 and S9 [53]). Furthermore, the convergence of $\mathbf{\Lambda}$ around 50 fs, followed by the convergence of spin around 100 fs, confirms the induction of spin from $\mathbf{\Lambda}$. Additionally, as illustrated in Fig. 4(c), we applied circularly polarized light to the bulk Pt to the *xy*-plane with the same intensity and frequency. The *z*-directional oscillating responses of the spin and $\mathbf{\Lambda}$ exhibit remarkably similar patterns of oscillations, whereas the orbital responses show a longer period [Fig. 4(d)]. Notably, these similarities are maintained regardless of the intensity or frequency of light, as shown in Figs. S10 and S11 [53].

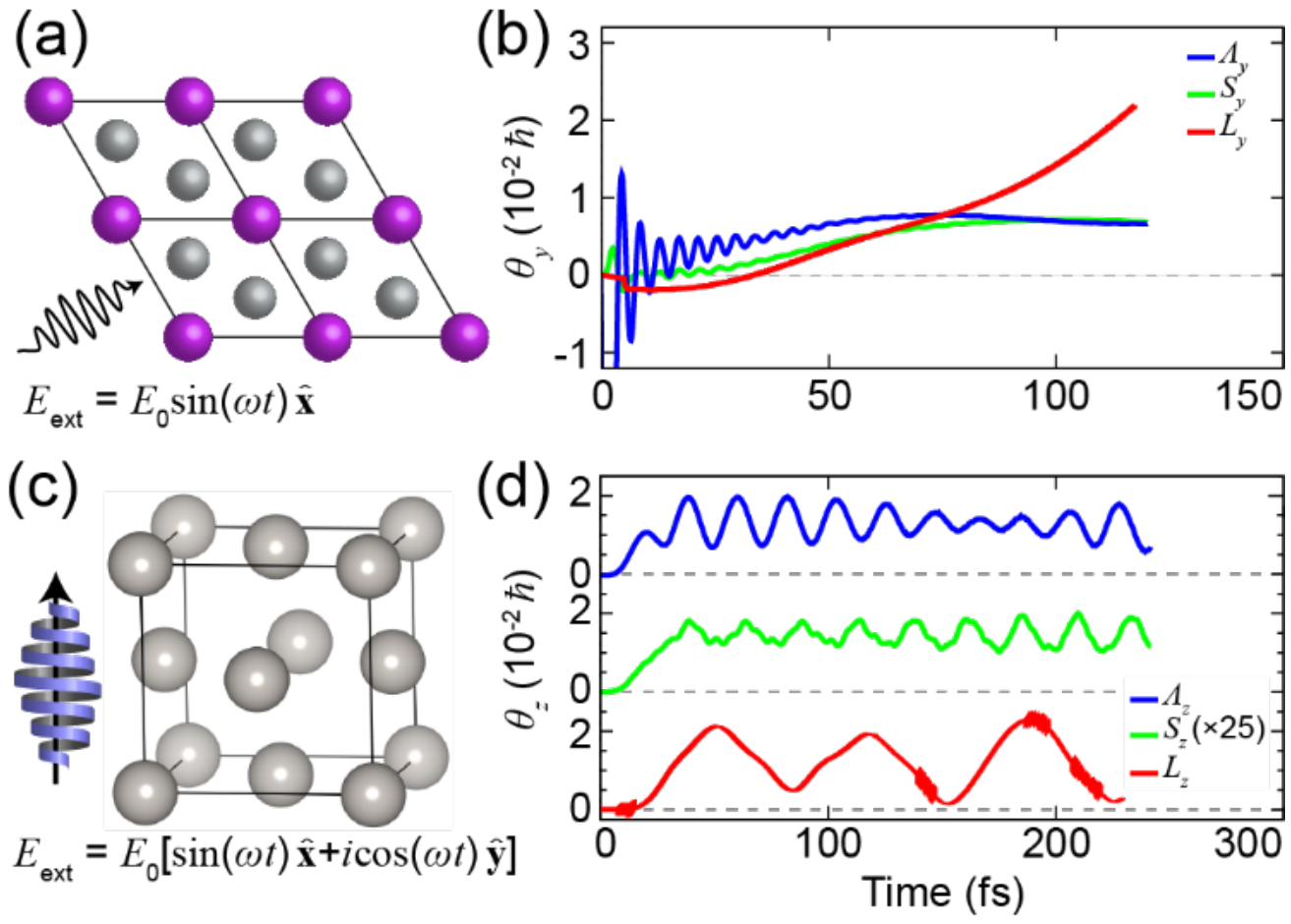

Fig. 4 (Color online) (a) Schematic illustration of linear-polarized light irradiated on a $BiAg_2$ monolayer. (b) Real-time profile of the $\mathbf{\Lambda}$ (blue), spin (green), and OAM (red) calculated by modern theory (red). The intensity and frequency of the incident light is $0.001\ \mathrm{V/Å}$ and frequency of $\hbar\omega = 1.0$ eV, respectively. (c) Schematic illustration of circular-polarized light irradiated on the fcc Pt. (d) Real-time profile of the $\mathbf{\Lambda}$ (blue), spin (green), and OAM (red). The intensity and frequency of the incident light is $0.001\ \mathrm{V/Å}$ and frequency of $\hbar\omega = 1.0$ eV, respectively.